\documentclass[%
 reprint,
superscriptaddress,
nofootinbib,
 amsmath,amssymb,
 aps,
]{revtex4-2}

\usepackage{graphicx}
\usepackage{dcolumn}
\usepackage{bm}
\usepackage{hyperref}
\usepackage{color}

\newcommand{\npipe}[0]{\texttt{NPIPE}}
\newcommand{\planck}[0]{\textit{Planck}}
\newcommand{\lb}[0]{\textit{LiteBIRD}}

\begin{document}


\title{Cosmic birefringence from \planck\ data release 4}

\author{P. Diego-Palazuelos}
\email{diegop@ifca.unican.es}
\affiliation{Instituto de F\'isica de Cantabria (CSIC-Universidad de Cantabria), Avda. de los Castros s/n, E-39005 Santander, Spain}
\affiliation{Dpto. de F\'isica Moderna, Universidad de Cantabria, Avda. de los Castros s/n, E-39005 Santander, Spain}
\author{J. R. Eskilt}
\email{j.r.eskilt@astro.uio.no}
\affiliation{Institute of Theoretical Astrophysics, University of Oslo, P.O. Box 1029 Blindern, N-0315 Oslo, Norway}
\author{Y. Minami}
\affiliation{Research Center for Nuclear Physics, Osaka University, Ibaraki, Osaka, 567-0047 Japan}
\author{M. Tristram}
\affiliation{Universit\'e Paris-Saclay, CNRS/IN2P3, IJCLab, 91405 Orsay, France}
\author{R. M. Sullivan}
\affiliation{Department of Physics \& Astronomy, University of British Columbia, 6224 Agricultural Road, Vancouver, British Columbia, Canada}
\author{A. J. Banday}
\affiliation{Universit\'e de Toulouse, UPS-OMP, IRAP, F-31028 Toulouse cedex 4, France}
\affiliation{CNRS, IRAP, 9 Av. colonel Roche, BP 44346, F-31028 Toulouse cedex 4, France}
\author{R. B. Barreiro}
\affiliation{Instituto de F\'isica de Cantabria (CSIC-Universidad de Cantabria), Avda. de los Castros s/n, E-39005 Santander, Spain}
\author{H. K. Eriksen}
\affiliation{Institute of Theoretical Astrophysics, University of Oslo, P.O. Box 1029 Blindern, N-0315 Oslo, Norway}
\author{K. M. G\'orski}
\affiliation{Jet Propulsion Laboratory, California Institute of Technology, 4800 Oak Grove Drive, Pasadena, California, U.S.A.}
\affiliation{Warsaw University Observatory, Aleje Ujazdowskie 4, 00-478 Warszawa, Poland}
\author{R. Keskitalo}
\affiliation{Computational Cosmology Center, Lawrence Berkeley National Laboratory, Berkeley, California, U.S.A.}
\affiliation{Space Sciences Laboratory, University of California, Berkeley, California, U.S.A.}
\author{E. Komatsu}
\affiliation{
Max Planck Institute for Astrophysics, Karl-Schwarzschild-Str. 1, D-85748 Garching, Germany
}
\affiliation{Kavli Institute for the Physics and Mathematics of the Universe (Kavli IPMU, WPI), Todai Institutes for Advanced Study, The University of Tokyo, Kashiwa 277-8583, Japan}
\author{E. Mart\'inez-Gonz\'alez}
\affiliation{Instituto de F\'isica de Cantabria (CSIC-Universidad de Cantabria), Avda. de los Castros s/n, E-39005 Santander, Spain}
\author{D. Scott}
\affiliation{Department of Physics \& Astronomy, University of British Columbia, 6224 Agricultural Road, Vancouver, British Columbia, Canada}
\author{P. Vielva}
\affiliation{Instituto de F\'isica de Cantabria (CSIC-Universidad de Cantabria), Avda. de los Castros s/n, E-39005 Santander, Spain}
\author{I. K. Wehus}
\affiliation{Institute of Theoretical Astrophysics, University of Oslo, P.O. Box 1029 Blindern, N-0315 Oslo, Norway}

\date{\today}
\begin{abstract}
We search for the signature of parity-violating physics in the cosmic microwave background, called cosmic birefringence, using the \planck\ data release 4. We initially find a birefringence angle of $\beta=0.30\pm0.11$ (68\% C.L.) for nearly full-sky data. The values of $\beta$ decrease as we enlarge the Galactic mask, which can be interpreted as the effect of polarized foreground emission. Two independent ways to model this effect are used to mitigate the systematic impact on $\beta$ for different sky fractions. We choose not to assign cosmological significance to the measured value of $\beta$ until we improve our knowledge of the foreground polarization.
\end{abstract}

\maketitle

\section{\label{sec:intro}Introduction}
Dark matter and dark energy in the Universe~\cite{Weinberg:2008zzc} may be a parity-violating pseudoscalar field, $\phi$, which changes sign under inversion of spatial coordinates~\cite{Marsh:2015xka,Ferreira:2020fam}.
This field can couple to the electromagnetic tensor $F_{\mu\nu}$ and its dual tensor $\tilde F^{\mu\nu}$ via a Chern-Simons term in the Lagrangian density, $\frac14g_{\phi\gamma}\phi F_{\mu\nu}\tilde F^{\mu\nu}$~\cite{Ni:1977zz,Turner:1987bw}, which makes the phase velocities of right- and left-handed states of photons different; thus, the plane of linear polarization rotates clockwise on the sky by an angle $\beta=-\frac12 g_{\phi\gamma}\int dt~\partial{\phi}/\partial t$, where $g_{\phi\gamma}$ is the coupling constant~\cite{Carroll:1989vb,Carroll:1991zs,Harari:1992ea,Carroll:1998zi}. The space filled with $\phi$ therefore behaves as if it were a birefringent material. For this reason, such an effect is often called ``cosmic birefringence.''

Linear polarization of the cosmic microwave background (CMB) photons is sensitive to $\beta$~\cite{Lue:1998mq}.
The polarization pattern on the sky can be decomposed into parity-even $E$ modes and parity-odd $B$ modes~\cite{Seljak:1996gy,Kamionkowski:1996zd}. The correlation functions of polarization fields (or the power spectra $C_\ell$ in spherical harmonics space with angular wavenumber $\ell$), contain two parity-even $EE$ and $BB$ auto-spectra, and one parity-odd $EB$ cross-spectrum.  Cosmic birefringence then yields $C_\ell^{EB,\mathrm{o}}=\frac12 \sin(4\beta)(C_\ell^{EE}-C_\ell^{BB})$, even when the intrinsic polarization contains no $EB$~\cite{Lue:1998mq,Feng:2004mq,Feng:2006dp,Liu:2006uh,Finelli:2008jv}. Here, the superscript ``o'' denotes the observed value, whereas the $EE$ and $BB$ power spectra on the right side are the ones before undergoing the cosmic birefringence. 

Recently, a weak signal of $\beta=0.35^\circ\pm 0.14^\circ$ (68\%~C.L.) was reported with a statistical significance of $2.4\,\sigma$~\cite{Minami:2020odp}, using analysis of $C_\ell^{EB}$ data from the European Space Agency \planck\ mission high-frequency instrument (HFI) public release~3 (PR3)~\cite{PlanckHFI:2018}.
Refs.~\cite{FujitaMinami:2020,Fujita:2020ecn,Takahashi:2020tqv,Mehta:2021pwf,Nakagawa:2021nme,Alvey:2021hjp,Choi:2021aze,Obata:2021nql} discuss possible cosmological implications of this particular measurement, and Refs.~\cite{Hinshaw:2012aka,Kaufman:2014rpa,Aghanim:2016fhp,Adachi:2019mjv,Bianchini:2020osu,Namikawa:2020ffr,Choi:2020ccd}
give previous constraints on $\beta$. In this paper, we apply the analysis method of Ref.~\cite{Minami:2020odp} to the \planck\ public release~4 (PR4), the so-called ``\npipe'' reprocessing of the \planck\ data~\cite{Akrami:2020bpw}, to further investigate the origin and robustness of this signal with respect to the known systematics in the \planck\ data sets and data cuts, as well as to the $EB$ correlation intrinsic to the polarized Galactic emission. 
Throughout this paper, we quote uncertainties at the 68\,\% confidence level (C.L.).

\section{\label{sec:data}The \npipe\ Data}
Details of the \npipe\ reprocessing of the \planck\ data are reported in Ref.~\cite{Akrami:2020bpw}.  Here, we briefly summarize the key aspects 
relevant to our analysis. The \npipe\ pipeline processes raw, uncalibrated detector data into polarized frequency and detector-set maps in the \texttt{HEALPix} format~\cite{Gorski:2004by}.  It is the only \planck\ pipeline that is designed to work with both low-frequency instrument (LFI)~\cite{PlanckLFI:2018} and HFI~\cite{PlanckHFI:2018} data.

The \npipe\ reprocessing achieved smaller noise by (1) including more data acquired during each 4-minute repointing maneuver (which makes up 9\,\% of the mission) between each of the 45-minute scans; and (2) better modeling the data via a short (167\,ms) baseline offset model for noise, suppressing degree-scale noise residuals.  HFI PR3 data were fitted with pointing period (between 35 and 75\,min long) offsets. A multi-frequency polarization model used in calibration greatly reduces large-scale polarization uncertainty but introduces a pipeline transfer-function that suppresses CMB polarization power at $\ell < 20$. Finally, a second-order analog-to-digital conversion nonlinearity (ADCNL) model is used.

The net effect of these differences is a scale-dependent reduction in the total uncertainty of $EE$ and $BB$: (1) About 50\,\% lower noise power spectrum ($N_\ell$) at $\ell \sim 10$; (2) 20--30\,\% lower $N_\ell$ at $\ell \sim 100$; and (3) 10--20\,\% lower $N_\ell$ at $\ell \sim 1000$ (this applies to temperature as well).

Much like the HFI PR3, \npipe\ fits and corrects for gain fluctuations, ADCNL, bolometric transfer-function residuals and bandpass mismatch by fitting time-domain templates while solving for the polarized map.    The release is accompanied by 600 total (signal, noise and systematics) simulations, each with realistic beam and pipeline transfer-functions and improved noise consistency.

\section{\label{sec:pipeline}Analysis method}
Our analysis pipeline is based upon Refs.~\cite{Minami:2019ruj,Minami:2020xfg,MinamiKomatsu:2020}, built independently by four groups (JRE, PDP, YM, MT): JRE, YM and MT follow the original implementation but with different $C_\ell$ estimation codes~\cite{Chon:2003gx,Alonso:2018jzx,Tristram:2004if}, while PDP uses an alternative implementation based on the small-angle approximation~\cite{delaHoz:2021vfx}.

If we relied only on $C_\ell^{EB}$ of the CMB, it would not be possible to distinguish between $\beta$ and miscalibration of the instrumental polarization angle, $\alpha$; thus, $\beta$ and $\alpha$ would be degenerate~\cite{Wu:2008qb,Miller:2009pt,Komatsu:2010fb,Keating:2012ge}. However, since $\beta$ is proportional to the path length of photons, the polarized emission from our Galaxy is only negligibly affected by $\beta$. We can use this property to break degeneracy between $\beta$ and $\alpha$~\cite{Minami:2019ruj}. Specifically, $C_\ell^{EB}$ of the CMB yields the sum $\alpha+\beta$, whereas foreground emission yields $\alpha$ in the absence of $EB$ intrinsic to the foreground.

We use four polarization-sensitive channels of the HFI at central frequencies of $\nu=(100, 143, 217, 353)$\,GHz. We split the detector sets on the focal plane into A and B sets following the definition given in Ref.~\cite{Akrami:2020bpw}. There are eight miscalibration angles, $\alpha_i$, with $i=100\textrm{A},\dots,353\textrm{B}$. We then exclude the auto-power spectra of the same maps, e.g., $100\textrm{A}\times 100\textrm{A}$, to avoid contamination of possible correlated noise. There are 28 unique pairs of A/B sets for $C_\ell^{EE}$ and $C_\ell^{BB}$, and 56 unique pairs for $C_\ell^{EB}$. 

We use a set of four Galactic masks removing respectively 5, 10, 20, and 30\,\% of the sky.
Each mask is constructed by thresholding the 353-GHz polarization and total intensity maps smoothed with a Gaussian with a full-width-at-half-maximum (FWHM) of $10^\circ$, to avoid polarized foreground residuals and potential intensity-to-polarization ($I\to P$) leakage residuals. 

We also exclude pixels in which the carbon monoxide (CO) line is bright. While CO is not polarized, the mismatch of detector bandpasses creates a spurious polarization signal via $I\to P$ leakage. We exclude pixels where CO is brighter than 45\,K$_\mathrm{RJ}$\,km\,s$^{-1}$. The Galactic and CO masks are apodized with a $1^\circ$ FWHM Gaussian taper.
While the CO strength varies over frequencies and there is no CO at 143\,GHz, we use a common CO mask for all frequencies to simplify the analysis because the CO mask removes less than 5\,\% of the sky, essentially on the Galactic plane where we expect the foreground model to fail anyway.
Finally, a common mask for point sources is constructed from the union of the \planck\ point-source masks at 100, 143, 217, and 353\,GHz.

The effective sky fraction in the covariance matrix of $C_\ell$ is given by $f_\mathrm{sky}=N_\mathrm{pix}^{-1}(\sum_{j=1}^{N_\mathrm{pix}} w_j^2)^2/(\sum_{j=1}^{N_\mathrm{pix}} w_j^4)$, where $w_i$ is the value of the (noninteger) 
apodized mask~\cite{Hivon:2002,Challinor:2004pr} and $N_\mathrm{pix}$ is the number of pixels. Our combined masks yield $f_\mathrm{sky}=0.93$, $0.90$, $0.85$, $0.75$, and $0.63$ for 0, 5, 10, 20, and 30\,\% Galactic masks, respectively.

We compute $C_\ell$ using \texttt{PolSpice} (JRE)
\cite{Chon:2003gx}\footnote{\url{http://www2.iap.fr/users/hivon/software/PolSpice/}}, \texttt{NaMaster} (PDP and YM)
\cite{Alonso:2018jzx}\footnote{\url{https://github.com/LSSTDESC/NaMaster}}, and \texttt{Xpol}
(MT) \cite{Tristram:2004if}\footnote{\url{https://gitlab.in2p3.fr/tristram/Xpol}}.
The $EB$ power spectra at low ($\ell\lesssim 10$) and high multipoles are sensitive to cosmic birefringence up to the epochs of reionization (a redshift of $z\approx 10$) and decoupling ($z\approx 1090$), respectively~\cite{Komatsu:2008hk,Sherwin:2021vgb}. We follow previous work~\cite{Aghanim:2016fhp,Minami:2020odp} and focus on the high-$\ell$ data, and bin $C_\ell$ from $\ell_\mathrm{min}=51$ to $\ell_\mathrm{max}=1490$ with a spacing of $\Delta\ell=20$. The number of bins is $N_\mathrm{bins}=72$. 

JRE, YM and MT use a Markov chain Monte Carlo to evaluate
$-2\ln L = \sum_{b=1}^{N_\mathrm{bins}}(\vec{v}^{\sf T}_b\mathbf{M}^{-1}_b\vec{v}_b +\ln|\mathbf{M}_b|)$,
and obtain the posterior distributions of $\beta$ and $\alpha_i$, while PDP minimizes $-2\ln L$ analytically in the small-angle approximation~\cite{delaHoz:2021vfx}.
Here, $b$ is the index for bins, and $\vec{v}_b\equiv \mathbf{A}\vec{C}_b^\mathrm{o}-\mathbf{B}\vec{C}_b^\mathrm{CMB,th}$ with $\vec{C}_b^\mathrm{o}=(C_b^{E_iE_j,\mathrm{o}}~C_b^{B_iB_j,\mathrm{o}}~C_b^{E_iB_j,\mathrm{o}})^{\sf T}$ and $\vec{C}_b^\mathrm{CMB, th}$ being the beam-smoothed and binned theoretical $\Lambda$ Cold Dark Matter ($\Lambda$CDM) best-fitting CMB spectra~\cite{planckcosmo:2018}.
The covariance matrix, $\mathbf{M}_b$, is binned from $\mathbf{M}_{\ell}= \mathbf{A}\mathrm{cov}(\vec{C}_{\ell}^\mathrm{o}, \vec{C}_{\ell}^\mathrm{o }{}^{\sf T})\mathbf{A}^{\sf T}$, and we divide $\mathbf{M}_b$ by $f_\mathrm{sky}$ to account for the mask.
Both $\mathbf{A}=\mathbf{A}(\alpha_i, \alpha_j)$ and $\mathbf{B}=\mathbf{B}(\alpha_i, \alpha_j, \beta)$ are block-diagonal matrices defined in Refs.~\cite{MinamiKomatsu:2020, Minami:2020odp}. 

\section{\label{sec:sim}The \npipe\ Simulation}

Not only $\alpha_i$, but also other instrumental systematics can create spurious $TB$ and $EB$ correlations~\cite{Miller:2009pt,Hivon:2016qyw}. To quantify the impact of the known systematics on $\beta$, we use realistic simulations of the \npipe\ processing, which include beam systematics, gain calibration and bandpass mismatches, ADCNL and the transfer-function correction, among others~\cite{Akrami:2020bpw}.

We use the A/B sets of CMB$+$noise$+$foreground realizations, with the foreground being the \texttt{Commander} sky model~\cite{PlanckComponentSeparation:2018}.
To isolate the systematics coupled to the CMB, we remove the beam-smoothed \texttt{Commander} maps from the simulations. The foreground-removed map still contains systematics coupled to the foreground, such as spurious polarization from the bandpass mismatch. 

Without foregrounds we can determine either $\alpha_i$ or $\beta$, assuming that the other vanishes. We find $\beta=-0.009^\circ\pm 0.003^\circ$ for $f_\mathrm{sky}=0.93$, where the uncertainty is that of the mean of 100 realizations. Therefore the impact of the known systematics on $\beta$ is negligible compared to the statistical uncertainty of the measurement, $0.11^\circ$, reported in Table~\ref{tab:result}. From the simulations we also find $\alpha_i=0.188^\circ\pm 0.009^\circ$, $-0.305^\circ\pm 0.007^\circ$, $0.047^\circ\pm 0.006^\circ$, $0.039^\circ\pm 0.005^\circ$, $-0.063^\circ\pm 0.008^\circ$, $0.020^\circ\pm 0.008^\circ$, $0.01^\circ\pm 0.03^\circ$, and $-0.06^\circ\pm 0.04^\circ$ for $i=100\textrm{A},\dots,353\textrm{B}$. These values do not need to agree with those of the data because the (unknown) miscalibration of the instrumental polarization angles and the cross-polarization response of beams are not included in the simulations. 

The $I\to P$ leakage gives $C_\ell^{EB}\propto C_\ell^{TT}$, whereas the cross-polarization effect gives $C_\ell^{EB}\propto C_\ell^{EE}$. We find that $C_\ell^{EB}$ of 100A and 100B resembles $C_\ell^{EE}$ of the CMB; thus, $\alpha_{100\textrm{A}}$ and $\alpha_{100\textrm{B}}$ detected in the simulations are due to the cross-polarization effect.
Because the impact on $\beta$ is small, we do not correct the values of $\alpha_i$ for these systematics.

\begin{table*}
	\caption{\label{tab:result}%
		Cosmic birefringence and miscalibration angles in units of degrees with $1\,\sigma~(68\,\%)$ uncertainties after correcting for the foreground $EB$ correlation. 
		The foreground $EB$ amplitudes, $A_\ell$, in four multipole bins are also shown.
	}
	\begin{ruledtabular}
		\begin{tabular}{cccccc}
			$f_\mathrm{sky}$ & 0.93 & 0.90 & 0.85 & 0.75 & 0.63 \\
			\colrule
			$\beta$  & $\phantom{-}0.36\pm 0.11$ & $\phantom{-}0.26 \pm 0.14$ & $\phantom{-}0.14\pm 0.17$ & $\phantom{-}0.10 \pm 0.21$ & $\phantom{-}0.29\pm 0.28$\\
			$\alpha_{100\textrm{A}}$ & $-0.32\pm 0.13$ & $-0.17\pm 0.16$ & $-0.07\pm 0.19$ & $-0.01\pm 0.23$ & $-0.21\pm 0.29$ \\
			$\alpha_{100\textrm{B}}$ & $-0.43\pm 0.13$ & $-0.32\pm 0.16$ & $-0.20\pm 0.19$ & $-0.14\pm 0.22$ & $-0.28\pm 0.29$ \\
			$\alpha_{143\textrm{A}}$ & $\phantom{-}0.03\pm 0.11$ & $\phantom{-}0.13\pm 0.14$ & $\phantom{-}0.29\pm 0.18$ & $\phantom{-}0.40\pm 0.21$ & $\phantom{-}0.22\pm 0.28$ \\
			$\alpha_{143\textrm{B}}$ & $\phantom{-}0.15\pm 0.11$ & $\phantom{-}0.25\pm 0.14$ & $\phantom{-}0.37\pm 0.18$ & $\phantom{-}0.39\pm 0.22$ & $\phantom{-}0.21\pm 0.28$ \\
			$\alpha_{217\textrm{A}}$ & $-0.06\pm 0.11$ & $\phantom{-}0.10\pm 0.14$ & $\phantom{-}0.22\pm 0.17$ & $\phantom{-}0.21\pm 0.21$ & $\phantom{-}0.02\pm 0.28$ \\
			$\alpha_{217\textrm{B}}$ & $-0.07\pm 0.11$ & $\phantom{-}0.07\pm 0.14$ & $\phantom{-}0.23\pm 0.17$ & $\phantom{-}0.23\pm 0.21$ & $0.003\pm 0.28$ \\
			$\alpha_{353\textrm{A}}$ & $-0.19\pm 0.10$ & $-0.08\pm 0.13$ & $\phantom{-}0.12\pm 0.17$ & $\phantom{-}0.03\pm 0.21$ & $-0.09\pm 0.28$ \\
			$\alpha_{353\textrm{B}}$ & $-0.23\pm 0.11$ & $-0.10\pm 0.13$ & $\phantom{-}0.10\pm 0.17$ & $\phantom{-}0.02\pm 0.21$ & $-0.02\pm 0.29$ \\
			$10^2A_{51-130}$ & $2.5^{+1.6}_{-1.4}$ & $5.7^{+2.5}_{-2.4}$ & $3.4^{+1.8}_{-1.7}$ & $18.8^{+6.0}_{-6.1}$ & $14.1^{+3.8}_{-3.7}$\\
			$10^2A_{131-210}$ & $0.8^{+1.2}_{-0.6}$ & $4.3^{+5.3}_{-3.1}$ & $9.8^{+4.2}_{-4.0}$ & $4.3^{+3.4}_{-2.8}$ & $2.6^{+2.9}_{-1.8}$ \\
			$10^2A_{211-510}$ & $1.5^{+2.4}_{-1.1}$ & $7.3^{+6.0}_{-4.7}$ & $5.1^{+4.9}_{-3.4}$ & $1.6^{+2.2}_{-1.2}$ & $3.1^{+3.2}_{-2.1}$ \\
			$10^2A_{511-1490}$ & $6.2^{+5.7}_{-4.1}$ & $4.2^{+4.2}_{-2.9}$ & $6.2^{+6.7}_{-4.3}$ & $4.9^{+5.3}_{-3.4}$ & $5.8^{+5.3}_{-3.8}$
		\end{tabular}
	\end{ruledtabular}
\end{table*}

\section{\label{sec:results}Results}
\begin{figure}
\centering
\includegraphics[width=0.95\linewidth]{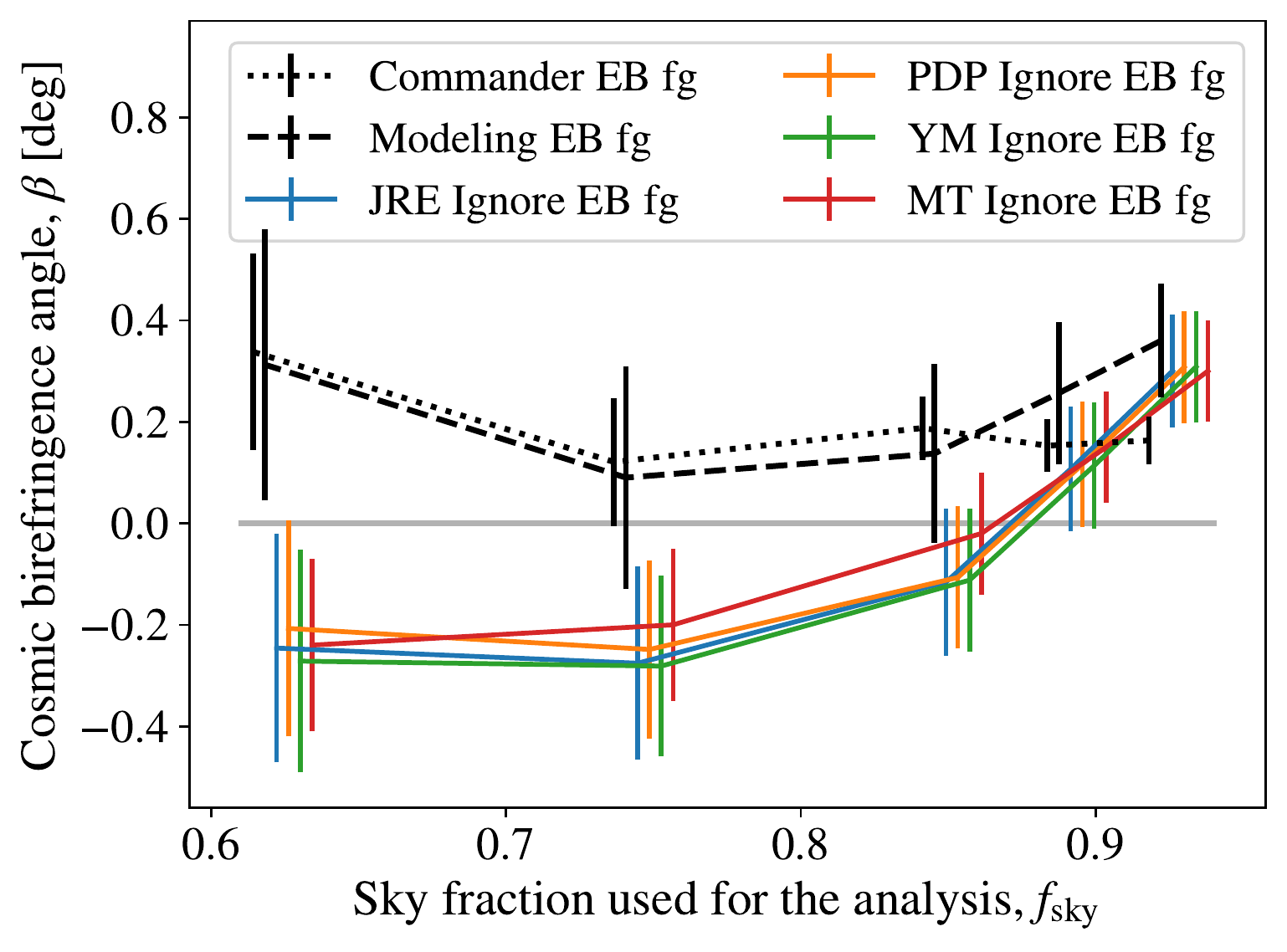}
\caption{\label{fig:beta-fsky}
Constraints on $\beta$ for various values of $f_\mathrm{sky}$ with and without accounting for the foreground $EB$ correlations.
For the former, the dashed and dotted lines show corrections using the filament model (Eq.~\eqref{eq:filament}) and the \texttt{Commander} sky model, respectively. For the latter, the results of four pipelines (JRE, PDP, YM, MT) are shown.}
\end{figure}

We show the measured values of $\beta$ in Fig.~\ref{fig:beta-fsky}. The results of the four pipelines agree, and we quote the results of the JRE pipeline throughout this paper, unless otherwise noted.
When we use all the data with $f_\mathrm{sky}=0.93$, we find $\beta= 0.30^\circ\pm 0.11^\circ$. 
When we remove one frequency channel at a time, we find
$\beta= 0.34^\circ\pm 0.12^\circ$ (without 100\,GHz), $0.38^\circ\pm 0.12^\circ$ (143\,GHz), $0.27^\circ\pm 0.13^\circ$ (217\,GHz), and $0.08^\circ\pm 0.21^\circ$ (353\,GHz). The cross-power spectra with 353\,GHz make the largest contribution.

We observe a decreasing $\beta$ for smaller $f_\mathrm{sky}$. This shift of $\beta$ goes along the degeneracy line of $\alpha_i+\beta=\mathrm{constant}$. A possible explanation for this trend is an $EB$ correlation intrinsic to the polarized dust emission~\cite{Minami:2019ruj,Minami:2020xfg,Minami:2020odp,Clark:2021kze}. When the foreground $EB$ power spectrum $C_\ell^{EB,\mathrm{FG}}$ exists, we have the relation $C_\ell^{EB,\mathrm{FG,o}}=\frac12\sin(4\alpha)(C_\ell^{EE,\mathrm{FG}}-C_\ell^{BB,\mathrm{FG}})+C_\ell^{EB,\mathrm{FG}}\cos(4\alpha)$ for a single channel~\cite{Abitbol:2015epq}. We can formally rewrite this as 
\begin{equation}
C_\ell^{EB,\mathrm{FG,o}}=\sqrt{J_\ell^2+\left(C_\ell^{EB,\mathrm{FG}}\right)^2}\sin(4\alpha+4\gamma_\ell)\,,
\end{equation}
where 
$J_{\ell}\equiv (C_\ell^{EE,\mathrm{FG}}-C_\ell^{BB,\mathrm{FG}})/2$ and $\tan(4\gamma_\ell)\equiv C_\ell^{EB,\mathrm{FG}}/J_\ell$. Here,
$\gamma_\ell$ is an effective angle for the foreground $EB$. For $C_\ell^{EB,\mathrm{FG}}\propto J_\ell$, $\gamma_\ell$ becomes independent of $\ell$, $\gamma_\ell=\gamma$, and is degenerate with $\alpha$. In this limit and $|\gamma|\ll 1$, the foreground does not yield $\alpha$ but $\alpha+\gamma$, and we measure
$\beta-\gamma$~\cite{Minami:2019ruj}. The sum, $\alpha+\beta$, is not affected. 

As both positive $TE$ and $TB$ correlations are found for polarized dust emission in the \planck\ data when the Galactic plane is masked~\cite{planckdust:2018,Weiland:2019uwg}, we expect $C_\ell^{EB,\mathrm{dust}}>0$ for the same mask~\cite{Huffenberger:2019mjx}. This is also confirmed by an independent analysis using the distribution of filaments of neutral hydrogen atoms~\cite{Clark:2021kze}.
From the ratio of $C_\ell^{TB}$ and $C_\ell^{TE}$ at 353\,GHz, $\gamma_\ell\approx \gamma>0$ is inferred for $\ell\lesssim 500$ when the Galactic plane is masked with $f_\mathrm{sky}\approx 0.7$~\cite{Clark:2021kze}. On the other hand, $\gamma_\ell$ oscillates around zero for nearly full-sky data. Therefore, we expect the observed $\beta$ (which measures some average of $\beta-\gamma_\ell$ over $\ell$) to decrease as we enlarge the Galactic mask, as observed. 

We estimate $\gamma=r^{EB,\mathrm{dust}}\sqrt{\xi}/[2(1-\xi)]$~\cite{Minami:2019ruj}, where $r^{EB,\mathrm{dust}}$ is the dust $EB$ cross-correlation coefficient and $\xi\equiv C_\ell^{BB,\mathrm{dust}}/C_\ell^{EE,\mathrm{dust}}\approx 0.5$~\cite{planckdust:2016}.
Ref.~\cite{Clark:2021kze} suggests that $r^{EB,\mathrm{dust}}= 4|r^{TB,\mathrm{dust}}|\psi^\mathrm{dust}$ with $r^{TB,\mathrm{dust}}\approx 0.05$~\cite{planckdust:2018} and $\psi^\mathrm{dust}\approx 5^\circ$ as a model for the signed \textit{upper bound} on $r^{EB,\mathrm{dust}}$ with $f_\mathrm{sky}\approx 0.7$. This gives $\gamma\lesssim 0.7^\circ$. The observed shift in $\beta$ from nearly full-sky data to $f_\mathrm{sky}\approx 0.7$ is $0.5^\circ$, consistent with the estimate.

\section{\label{sec:ebdust}Modeling the impact of the foreground \texorpdfstring{$EB$}{EB} correlation}
We apply two independent approaches to assess the foreground impact.
The first model for $C_\ell^{EB,\mathrm{dust}}$
is based on filaments of hydrogen clouds producing the thermal dust emission and polarization~\cite{Huffenberger:2019mjx,Clark:2021kze}. We use this model because it is the only physical model for the foreground $EB$ available today. When the filaments and the magnetic field lines are perfectly aligned, the model produces a positive $TE$ but no $TB$ or $EB$ correlations. When they misalign by a small angle $\psi^\mathrm{dust}$, $TB$ and $EB$ correlations emerge with the same sign. We thus use the ansatz
$C_\ell^{EB,\mathrm{dust}}
    = A_\ell C_\ell^{EE,\mathrm{dust}}\sin(4\psi_\ell^\mathrm{dust})$ with $0\le A_\ell\ll 1$ and $\psi_\ell^\mathrm{dust}=\frac12\arctan(C_\ell^{TB,\mathrm{dust}}/C_\ell^{TE,\mathrm{dust}})$. 
In the model of Ref.~\cite{Clark:2021kze} $A_\ell\to |r_\ell^{TB,\mathrm{dust}}|$ and $C_\ell^{EE,\mathrm{dust}}\to \sqrt{C_\ell^{EE,\mathrm{dust}}C_\ell^{BB,\mathrm{dust}}}$; however, we treat $A_\ell$ as a free parameter here, and use $C_\ell^{EE,\mathrm{dust}}$ because the data suggest $C_\ell^{EE,\mathrm{dust}}\propto C_\ell^{BB,\mathrm{dust}}$ and taking the square-root of noisy data is not numerically stable.
When the angles are small, we obtain 
\begin{equation}
\label{eq:filament}
  \gamma_\ell\simeq 
\frac{A_\ell C_\ell^{EE,\mathrm{dust}}}{C_\ell^{EE,\mathrm{dust}}-C_\ell^{BB,\mathrm{dust}}}\frac{C_\ell^{TB,\mathrm{dust}}}{C_\ell^{TE,\mathrm{dust}}}\,.
\end{equation}
We use this model to account for the possible impact of the dust $EB$ correlation. 

We specifically modify $\mathbf{A}$ and $\mathbf{B}$ in $-2\ln L$ as $\mathbf{A}= \left(
\begin{array}{cc}-\vec{\Lambda}^{\sf T}_{\ell}\mathbf{\Lambda}_{\ell}^{-1}, 1\end{array}\right)$ and
\begin{equation}
\mathbf{B}=\left(
\begin{array}{cc}\vec{R}^{\sf T}(\alpha_i+\beta, \alpha_j+\beta)-\vec{\Lambda}^{\sf T}_\ell\mathbf{\Lambda}_\ell^{-1}\mathbf{R}(\alpha_i+\beta, \alpha_j+\beta)\end{array}\right)\,,
\end{equation}
where $\mathbf{R}$ and $\vec{R}$ are defined in equations~8 and 9 of Ref.~\cite{MinamiKomatsu:2020}, respectively. We define the new matrices,
${\bf\Lambda}_\ell=\mathbf{R}(\alpha_i, \alpha_j)+\mathbf{D}(\alpha_i, \alpha_j)\mathbf{F}_\ell$ and $\vec{\Lambda}_\ell^{\sf T}=\vec{R}^{\sf T}(\alpha_i, \alpha_j)+\vec{D}^{\sf T}(\alpha_i, \alpha_j)\mathbf{F}_\ell$, where
\begin{eqnarray}
\bf{D}&=&\left(
\begin{array}{cc}
-\cos(2\alpha_i)\sin(2\alpha_j)&
-\sin(2\alpha_i)\cos(2\alpha_j)\\
\sin(2\alpha_i)\cos(2\alpha_j)&
\cos(2\alpha_i)\sin(2\alpha_j)
\end{array}\right)\,,
\end{eqnarray}
\begin{equation}
\vec{D}=\left(
\begin{array}{c}
\cos(2\alpha_i)\cos(2\alpha_j)\\
-\sin(2\alpha_i)\sin(2\alpha_j)
\end{array}\right)\,,\quad
\mathbf{F}_\ell=\left(
\begin{array}{cc}
2\gamma_{\ell,j}&-2\gamma_{\ell,j}\\
2\gamma_{\ell,i}&-2\gamma_{\ell,i}
\end{array}
\right)\,.
\end{equation}
We retain the possibility that $\gamma_{\ell,i}$ may depend on the $i$th frequency channel. 

We compute the ratios of $C_\ell^{XY,\mathrm{dust}}$ in Eq.~\eqref{eq:filament} using the 353\,GHz data. 
To reduce the scatter, we smooth $C_\ell^{XY}$ by applying a one-dimensional Gaussian filter before computing the ratios. Since the dust spectral energy distribution (SED) cancels, these ratios can be used to fit the cross-power spectra of all the other frequency channels. In this model, we expect $A_\ell$, hence $\gamma_\ell$, to be independent of frequencies. We confirm that allowing for frequency-dependent $A_\ell$ yields similar results; thus, we use the same $A_\ell$ for all the frequency combinations. To account for possible dependence on $\ell$, we split $A_\ell$ into four bins ($51\leq \ell\leq 130$, $131\leq\ell\leq210$, $211\leq\ell\leq510$ and $511\leq\ell\leq1490$). 

In Fig.~\ref{fig:beta-fsky}, we show $\beta$ from the simultaneous fit of $\alpha_i$, $\beta$, and $\gamma_\ell$. The model brings $\beta$ back to positive values for all $f_\mathrm{sky}$, confirming that the declining $\beta$ is caused by $C_\ell^{EB,\mathrm{dust}}$. We report the numerical values in Table~\ref{tab:result}.

We also try another, completely different approach  to assessing the foreground impact using the \npipe\ simulation of the \texttt{Commander} sky model~\cite{PlanckComponentSeparation:2018} based on power-law synchrotron and one-component modified blackbody (MBB) dust SED. 
When the angular resolution of the foreground model is higher than that of the target frequency channel, the foreground component is smoothed to match the \texttt{QuickPol}~\cite{Hivon:2016qyw} beam specific to the \npipe\ data set. To avoid divergence in the deconvolution, the Gaussian beam of $5'$ FWHM present in \texttt{Commander}'s dust component is maintained at 217 and 353\,GHz. We measure $C_\ell^{EB,\mathrm{FG}}$ from the simulations and include it in the PDP pipeline with a free amplitude ${\cal D}$, which is fit simultaneously with $\beta$ and $\alpha_i$~\cite{Diego-Palazuelos}.

The measured $C_\ell^{EB,\mathrm{FG}}$ does not represent a signal-dominated template for the foreground $EB$, but is
dominated by the statistical fluctuation with the variance proportional to $C_\ell^{EE,\mathrm{FG}}C_\ell^{BB,\mathrm{FG}}/(2\ell+1)$. Thus, using this in the fit requires great care; since the \texttt{Commander} sky model was taken from the data, $C_\ell^{EB,\mathrm{FG}}$ fluctuates in the same way as the data, resulting in surprisingly tight constraints on the parameters. We find amplitude parameters of ${\cal D}=1.03\pm 0.01$, $1.02\pm 0.01$, $1.03\pm 0.01$, $1.01\pm 0.04$, and $1.13\pm 0.05$ in descending order of $f_\mathrm{sky}$. This does \textit{not} mean that we detect the foreground $EB$ with high statistical significance, but merely shows that the data are consistent with $C_\ell^{EB,\mathrm{FG}}$ taken from the simulation on a mode by mode basis.

In Fig.~\ref{fig:beta-fsky}, we show that the behavior of $\beta$ is qualitatively similar to that from the filament model. The values are $\beta= 0.16^\circ\pm 0.05^\circ$, $0.15^\circ\pm 0.05^\circ$, $0.19^\circ\pm 0.06^\circ$, $0.12^\circ\pm 0.13^\circ$, and $0.34^\circ\pm 0.19^\circ$ in descending order of $f_\mathrm{sky}$, which agree well with those of the filament model except for $f_\mathrm{sky}=0.93$. This discrepancy could be due to the complexity of the foreground emission near the Galactic plane that is not captured by the power-law synchrotron and MBB dust SED in the \texttt{Commander} model. The derived uncertainties in $\beta$ are smaller because $C_\ell^{EB,\mathrm{FG}}$ is correlated well with the data, including the statistical fluctuations, hence reducing (somewhat artificially) the covariance in the likelihood. 

Let us comment on another limitation of this approach. The existence of $\alpha_i$ inevitably yields a spurious $C_\ell^{EB,\mathrm{FG}}$ in the \texttt{Commander} map. If $\alpha_i$ varies over frequency channels $i$, the foreground SED of polarization fields in the sky assumed for \texttt{Commander} is no longer adequate for describing the data, further yielding a spurious $EB$. Therefore we must be careful when interpreting $C_\ell^{EB,\mathrm{FG}}$ in the \texttt{Commander} map. The modeling approach (Eq.~\eqref{eq:filament}) does not suffer from this issue, but the model itself may be limited. These two approaches are complementary, and it is encouraging that they yield similar results. We adopt $\beta$ from the filament model as our baseline results.

\section{\label{sec:conclusion}Conclusions}
In this paper, we searched for the signal of cosmic birefringence in the {\it Planck} PR4 data. First, we applied the methodology of Ref.~\cite{Minami:2020odp}, finding $\beta=0.30^\circ\pm 0.11^\circ$ for nearly full-sky data. This agrees with, and is more precise than, the previous estimate from the PR3 data set~\cite{Minami:2020odp}. 

We then expanded the methodology and explored the dependence of $\beta$ on Galactic masks. We found a trend of decreasing $\beta$ for smaller $f_\mathrm{sky}$, which can be understood as the effect of $C_\ell^{EB,\mathrm{dust}}$~\cite{Minami:2019ruj,Minami:2020xfg,Minami:2020odp,Clark:2021kze}.
Accounting for this effect in two independent ways, we found that $\beta$ was positive for all $f_\mathrm{sky}$.
Which $f_\mathrm{sky}$ value should we choose for the most robust determination of $\beta$? We cannot provide a definitive answer until we improve our understanding of the foreground $EB$.

If confirmed as a cosmological signal, this would provide evidence for physics beyond the standard model of elementary particles and fields, with profound implications for fundamental physics. To make progress, we must search for $\beta$ in independent data sets of on-going~\cite{Xu:2019rne,Polarbear:2020lii,ACT:2020gnv,Dutcher:2021vtw,BICEPKeck:2021gln,Ade:2021cyk} and future~\cite{Ben:2018,Hui:2018cvg,Ade:2018sbj,abazajian2019cmbs4,Hazumi2019} experiments. Since the largest statistical significance is seen for nearly full-sky data, a full-sky mission such as \lb~\cite{Hazumi2019} will play an important role. 

An overarching goal of many planned CMB experiments is to find the signature of primordial gravitational waves in $B$-mode polarization~\cite{Kamionkowski:2015yta}, which drives requirements for experimental design. The $EB$ science may provide additional requirements for, e.g., cross-polarization coupling and beam systematics. Equally important is the need for high-fidelity end-to-end simulations, with the $EB$ science in mind.

\begin{acknowledgments}
We thank D.~J. Watts, S. Clark and J.~C. Hill for useful discussions and B. Partridge for comments on the draft. \planck\ is a project of the European Space Agency (ESA) with instruments provided by two scientific consortia funded by ESA member states and led by Principal Investigators from France and Italy, telescope reflectors provided through a collaboration between ESA and a scientific consortium led and funded by Denmark, and additional contributions from NASA (USA). Some of the results in this paper have been derived using the \texttt{HEALPix} package. This research used resources of the National Energy Research Scientific Computing Center (NERSC), a U.S. Department of Energy Office of Science User Facility operated under Contract No.~DE-AC02-05CH11231. Part of the research was carried out at the Jet Propulsion Laboratory, California Institute of Technology, under a contract with the National Aeronautics and Space Administration (80NM0018D0004). JRE, HKE and IW acknowledge funding from the European Research Council (ERC) under the Horizon 2020 Research and Innovation Programme (Grant agreement No.~819478).
The work of YM was supported in part by the Japan Society for the Promotion of Science (JSPS) KAKENHI, Grants No.~JP20K14497. RS and DS acknowledge the support of the Natural Sciences and Engineering Research Council of Canada.
PDP acknowledges financial support from the \textit{Formaci\'on del Profesorado Universitario (FPU) Programme} of the Spanish Ministerio de Ciencia, Innovaci\'on y Universidades, and thanks the Max Planck Institute for Astrophysics for hospitality when this work was being finalized. PDP, BB, EMG and PV thank the Spanish Agencia Estatal de Investigaci\'on (AEI, MICIU) for the financial support provided under the projects with references PID2019-110610RB-C21, ESP2017-83921-C2-1-R and AYA2017-90675-REDC, co-funded with EU FEDER funds, and acknowledge supports from Universidad de Cantabria and Consejer{\'i}a de Universidades, Igualdad, Cultura y Deporte del Gobierno de Cantabria via the “Instrumentaci{\'o}n y ciencia de datos para sondear la naturaleza del universo” project, as well as from Unidad de Excelencia Mar{\'i}a de Maeztu (MDM-2017-0765).
The work of EK was supported in part by JSPS KAKENHI Grant No.~JP20H05850 and JP20H05859, and the Deutsche Forschungsgemeinschaft (DFG, German Research Foundation) under Germany's Excellence Strategy - EXC-2094 - 390783311.
The Kavli IPMU is supported by World Premier International Research Center Initiative (WPI), MEXT, Japan.
\end{acknowledgments}
\bibliographystyle{apsrev4-2}
\bibliography{references}
\end{document}